# Efficient Deep Neural Network Accelerator Using Controlled Ferroelectric Domain Dynamics

*Sayani Majumdar\**

S. Majumdar

VTT Technical Research Centre of Finland Ltd., P.O. Box 1000, FI-02044 VTT, Finland.

E-mail: sayani.majumdar@vtt.fi

**Abstract**. The current work reports an efficient deep neural network (DNN) accelerator where analog synaptic weight elements are controlled by ferroelectric domain dynamics. An integrated device-to-algorithm framework for benchmarking novel synaptic devices is used. In P(VDF-TrFE) based ferroelectric tunnel junctions, analog conductance states are measured using a custom pulsing protocol and associated control circuits and array architectures for DNN training is simulated. Our results show precise control of polarization switching dynamics in multi-domain, polycrystalline ferroelectric thin films can produce considerable weight update linearity in metal-ferroelectric-semiconductor (MFS) tunnel junctions. Ultrafast switching and low junction current in these devices offer extremely energy efficient operation. Through an integrated platform of hardware development, characterization and modelling, we predict the available conductance range where linearity is expected under identical potentiating and depressing pulses for efficient DNN training and inference tasks. As an example, an analog crossbar based DNN accelerator with MFS junctions as synaptic weight elements showed > 93% training accuracy on large MNIST handwritten digit dataset while for cropped images, a > 95% accuracy is achieved. One observed challenge is rather limited dynamic conductance range while operating under identical potentiating and depressing pulses below 1V.

Investigation is underway for improving the FTJ dynamic conductance range maintaining the weight update linearity under identical pulse scheme.

## 1. Introduction

Today's data-driven world led to intense research effort towards alternate computing paradigm beyond the von Neumann architecture. The current computing paradigms are suitable for precise scientific computation but clearly becoming insufficient for real-time processing of large amount of unstructured data and unsustainable in terms of energy and environmental footprint. Neuromorphic computing, inspired by human brain like parallel and in-memory data processing, emerged as one viable choice in handling raw sensory data already at the sensor node. [1, 2] This includes feature extraction, classification, error correction, decision-making, event-based communication, etc. [3, 4] In biological brains, learning takes place through synaptic weight change in response to incoming electrical impulses from the pre and post-synaptic neurons. Electric impulse driven release and reception of ions by the neurons cause the synaptic weight change, as shown in Figure 1(a). Gradual change in synaptic weight is the key to our energy efficient learning and processing ability.

Two-terminal memory resistors or "memristors" having multiple, reproducible intermediate conductance states in addition to their on and off states are promising candidates as electronic synapses. Momentary pulsed excitation in memristive synapses lead to various synaptic plasticity features making them suitable for learning and adapting to input signals.[5] Different types of memristors, like, conductive filament memories,[6,7] phase change memories,[8] spin-based memories,[9] and ferroelectric (FE) memories [10] have been proposed as high performance synapses for their properties like nonvolatile and gradual conductance change, a threshold feature, a simple structure for dense integration and energy efficiency.[11]

Among different memory technologies, ferroelectric field driven devices can offer unique functionalities and a vast, new design space due to programmable and stable up and down polarization states of the ferroelectric thin films, controllable intermediate states in multi-domain FE structures and ultralow off and standby power. Among different ferroelectric memory technologies, ferroelectric random access memories (FeRAMs), [12] composed of one transistor one capacitor structures (1T1C), is the more matured technology. In FeRAMs, with a ferroelectric capacitor as the main charge storage component, however, the possibility of dense integration gets restricted. Additionally, readout in these devices are destructive and frequent reprogramming makes the operation cost higher. [2] In recent years, ferroelectric field effect transistors (FeFETs) [13, 14] and ferroelectric finFETs (Fe-finFETs) [15, 16] have emerged as densely integrable analog memories and synaptic devices with fabrication processes that are compatible to standard semiconductor technologies. This renewed the drive towards finding high performance ferroelectric devices for analog memories and neuromorphic computing. Among this list, ferroelectric tunnel junctions (FTJs) have emerged as one of the potential analog resistive memories due to their simple device structure and voltage-controlled ultrafast operation leading to extremely energy efficient and non-volatile, multibit data retention. [10, 11, 17] In an FTJ, an ultrathin ferroelectric film is sandwiched between two dissimilar metal electrodes giving rise to an asymmetric barrier potential profile. Due to an applied electric field, the net polarization of the FE layer rotates between two perpendicular directions and based on the polarization direction, charge carriers at the metal electrodes either are attracted to or repelled from the FE-metal interface. Accumulation of carriers result in lower barrier height for the tunneling electrons while that for the depletion result in higher barrier height for the tunneling electrons, leading to device on and off states. In multi-domain ferroelectrics, it is possible to control the ratio of up and down polarized domains in a way that gives rise to multiple, stable conductance states instead of only on and off, making them suitable for analog memory application and as programmable synaptic weight elements for deep neural networks (DNN) training and inference.

Reported results on conventional perovskite FE based FTJs mostly concentrate on binary switching properties. [10, 18-20] However, in recent times, experimental results showed that utilizing the mixed polarization phase of the FE ultrathin films, it is possible to attain multiple conductance states in FTJs. [21-24] Boyn et al. showed FTJs are able to perform unsupervised spike-timing-dependent plasticity (STDP) learning involving controlled ferroelectric domain switching. Custom-designed pulse shapes were used to obtain desired STDP responses. [25] In our previous works, we demonstrated that organic FTJs are able to replicate all synaptic plasticity features and STDP learning and all within ultrafast timescales. [20, 22-24] However, their system level performance in network training experiments were not evaluated.

In the field of artificial intelligence, DNNs demonstrated successful performance in different cognitive tasks like speech and image recognition. [26, 27] However, the energy consumption and training time of DNNs face challenges by off-chip memory access bottleneck due to the large memory requirements of the weight matrices. Therefore, it is predicted that for a fully connected DNN, an on-chip storage and weight updates at the same node, with all the nodes connected together in an array, can result in much reduced data trafficking. This can cause significant acceleration in training and improvement in training energy. For perovskite oxide based FTJs, [28] $Hf_{0.5}Zr_{0.5}O_2$ (HZO) based FTJs, [29] FeFETs, [13] Fe-thin film transistors (FeTFTs) and Fe-finFETs, [15, 16] efficient DNN training has been demonstrated. The FeFETs, Fe-finFETs and FeTFTs (one-transistor memory) have a larger dynamic range (On/Off ratio exceeding $10^6$ or similar) to operate on, which improves the training accuracy. However, in all three terminal devices, non-identical pulsing schemes have been used to achieve weight update linearity. Non-identical pulsing schemes bring additional circuit complexity. Also, three terminal devices in a crossbar configuration causes higher intricacies of the circuit and interconnect designs for full HW implementation of dense arrays. Additionally, due to higher number of mask layers, the device fabrication complexity, time and cost is higher for FeFETs and Fe-finFETs.

In the current work, we have utilized an organic metal-FE-semiconductor (MFS) tunnel junction and controlled the FE domain dynamics precisely in them with identical programming and erasing pulses of <1V amplitude. Based on the weight update performance, an FTJ based analog accelerator is simulated that showed > 93% training accuracy for DNN training on large MNIST handwritten digit dataset. For small MNIST handwritten digit dataset, the training accuracy exceeds 95%, approaching the ideal numerical simulation value of 98%. Future research directions are discussed for improving the dynamic conductance range, analog bit precision and hence the training accuracies without additional circuit or algorithmic complexity. Finally the results are benchmarked against state-of-the-art FE memory devices and analog accelerator performances based on these devices.

Here, it is important to mention that although all experimental data in the article is from Au/ Poly(vinylidene fluoride-trifluoroethylene), P(VDF-TrFE)/Nb-SrTiO$_3$ (NSTO) based FTJs, the device physics of nucleation limited switching (NLS) of FE domains is applicable to all polycrystalline ferroelectrics, and hence it can be generalized. The device-to-algorithm framework is even more general and can be applied for evaluating the performance of any analog memory based accelerator design.

## 2. Results and Discussion

**2.1. Analog conductance with controllable ferroelectric domain dynamics**

*2.1.1. Full and partial ferroelectric polarization reversal in FTJs*

Permanent electric dipole moments in P(VDF-TrFE), a long-chain copolymer with chemical formula [(CH$_2$-CF$_2$)$_n$–(CF$_2$-CHF)$_n$], point from electronegative fluorine to electropositive hydrogen atoms (Figure 1c). Ordering of the polymer chains into a parallel quasi-hexagonal close-packing structure gives rise to a macroscopic polarization of $P$ = 10 μC/cm$^2$ in the ferroelectric *β*-phase. [31] Polarization reversal in P(VDF-TrFE) takes place under electric field by rotation of the $CH_2$ and $CF_2$ molecular units around the backbone of the polymer chain. The polarization reversal mechanism has been explained in literature mostly on the basis of nucleation and growth model of polarized domains.[32,33] The domain nucleation and growth in the copolymer films can be significantly affected by the microstructure of the FE film, for instance, increased structural defects can hinder easy rotation of the molecular chains leading to domain pinning sites.[23] Depending on the amplitude and duration of an applied electric field ($E$) on a FE thin film, the net polarization of the FE film might or might not be fully rotated reaching the saturation value ($P_S$). If $E$ is large enough, i.e., above the coercive field ($E_C$) of the material, the FE polarization reaches saturation and upon withdrawal of $E$, shows an open hysteresis loop ($P$-$E$), known as a FE major loop. However, if $E$ is not sufficient to saturate the net polarization, the opening of the $P$-$E$ loop narrows, resulting in a FE minor loop. In memory devices, in addition to the magnitude $E$, the applied pulse width ($w$) is also of vital important since the shorter pulses require higher $E$ strength to switch the FE domains. The minor loop operations of FE memories primarily determine the analog bit precision and linearity of weight update. In multi-domain FE devices (both in FTJs and FE field-effect transistors, FeFETs), intermediate conductance states can be controlled by modifying FE film structural properties [23], FE film thickness, [20,23] electrode materials [20] and a custom-designed pulse programming scheme. [22,23] Longer retention of FE polarization depends on the proper screening of the polarization charges at the electrode-FE interface and is crucial for non-volatile data retention properties. Also for inference only networks, non-volatile retention of training data is of vital importance. In ultrathin ferroelectric films, the depolarization field is often strong and leads to faster loss of polarization. In order to verify this in our ultrathin ferroelectric film, we studied the long-term retention of FE polarization using Kelvin-probe-force microscope (KPFM) data (Figure 1(d-f)). For KPFM measurements, the FE films were first poled using the AFM tip in contact mode followed by non-contact probing of the surface potential of the films due to the bound polarization charges on the FE. A dc bias was applied on the tip while the NSTO substrate was grounded. The measured surface potential from KPFM measurements confirm that the polarized domains are stable over a period of 3 hours with only a little dispersion of surface polarization charges. This confirms the suitability of our FTJs for non-volatile data retention. To confirm the nature of ferroelectric switching in ultrathin P(VDF-TrFE) films, we also performed piezo-force microscopy (PFM) on the samples. Results for local PFM hysteresis of 6-nm-thick P(VDF-TrFE) films on NSTO are shown in Figure 1(g,h). Ultrathin P(VDF-TrFE) films show robust ferroelectricity, verified by near 180° phase contrast and butterfly-shaped amplitude curves. Our previous results confirm that the polarization switching and coercive fields of P(VDF-TrFE) films depend on the crystallinity, size and orientation of the grains and therefore can be utilized for linear conductance modulation by modifying the film microstructural properties. [23]

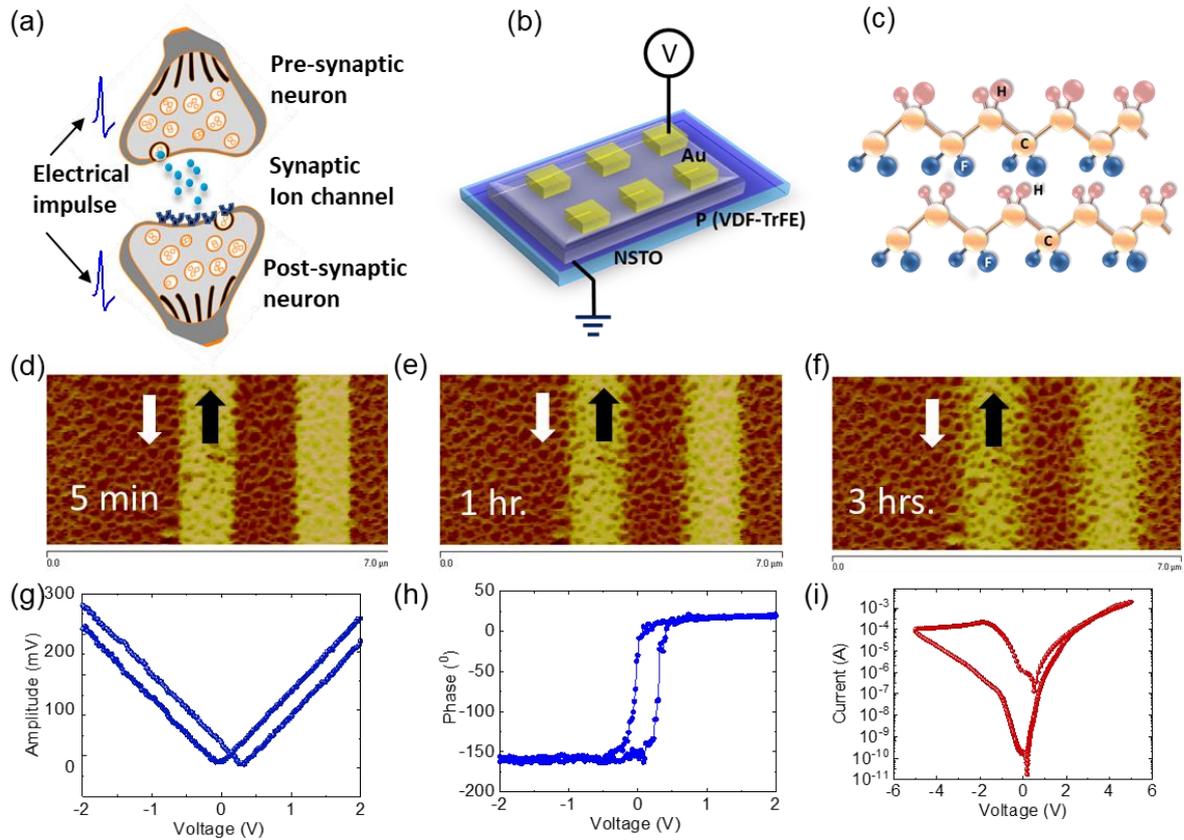

**Figure 1.** A schematic representation and electrical characterizations of the P(VDF-TrFE) based FTJ synaptic devices. a) A sketch of a biological synaptic system. b) A schematic illustration of a two-terminal Au/P(VDF-TrFE)/ NSTO FTJ device in which the Au and NSTO electrodes mimic the pre- and post- synapse, respectively. c) Molecular structure of the FE P(VDF-TrFE) copolymer. KPFM surface potential of P(VDF-TrFE) thin films after poling the domains with an applied voltage of 5V, d) immediately after poling, e) 1 hour after poling and f) 3 hours after poling showing long-term stability of the poled domains. g) PFM out-of-plane amplitude and h) phase of P(VDF-TrFE)/ NSTO films when the tip voltage was swept between ±2V. i) Current-voltage loops of the junctions under continuous bias sweep.

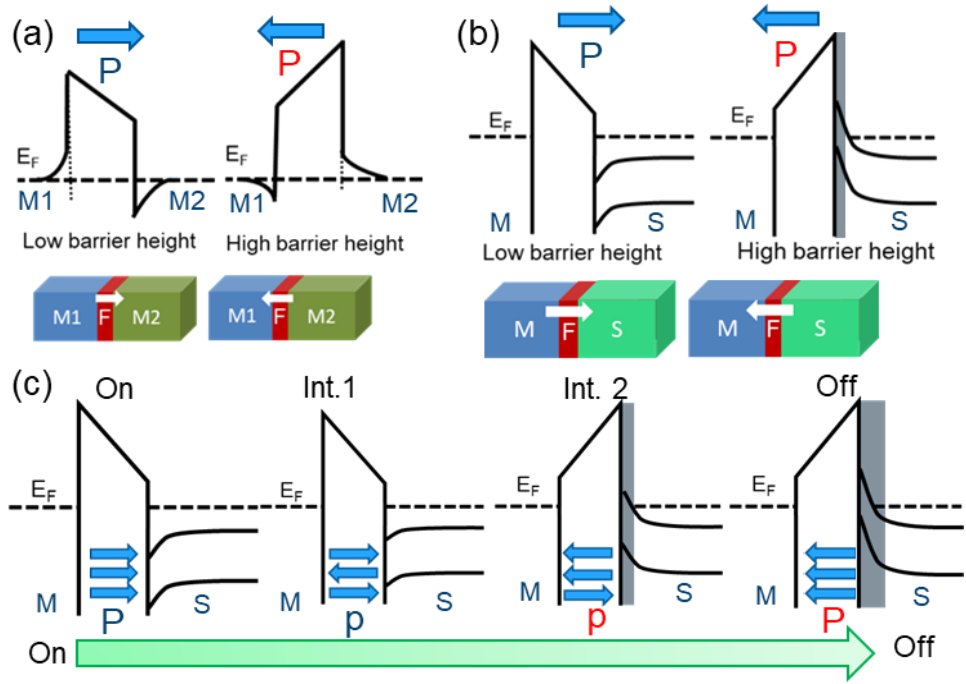

**Figure 2.** Schematic band diagram of the ferroelectric tunnel junctions with (a) metal-ferrolectric-metal (MFM) and (b) metal-ferroelectric-semiconductor (MFS) structures. The tunneling barrier profile of the FTJ at up and down polarization states of the FE polarization switching for both structures. (c) For an MFS structure, evolution of the barrier profile while sweeping the device from on to off states and reverse showing existence of different intermediate states due to partial polarization switching in the FE layer (Int.1 and Int.2) between two perpendicular polarization directions.

The reversal of FE polarization in FTJs leads to an open hysteresis loop in the current-voltage characteristics (Figure 1(i)) due to modified tunneling barrier height and width in certain cases, as discussed in the following paragraphs, making them suitable as two terminal memristive devices with non-volatile conductance states. In FTJs with two metal electrodes (Figure 2(a)), FE polarization reversal results in modified tunnel barrier height only and the resulting limitation appears in relatively low device on/off ratio. This could be overcome with utilization of one semiconductor electrode instead of two metal ones. Carrier depletion in a semiconductor electrode cause creation of a Schottky barrier that modifies the tunnel barrier width in FTJs in addition to the barrier height (Figure 2(b)), leading to a large dynamic conductance range. Multiple conductance states in these FTJs appear from partial polarization rotation induced intermediate barrier height and width modulation which can be controlled using FE film thickness, film morphology and $E$ modulation (Figure 2(c)). This has been shown experimentally in previous works, both for traditional oxide FTJs [19, 34] and from polymeric FTJs. [20, 22-24] In the present work, we used both experimental and integrated modelling framework to investigate the desired tuning conditions of the MFS junctions for multiple stable conductance states through precise control over partial polarization rotation of the ultrathin FE films.

For a thorough understanding of the switching dynamics, FTJ switching responses were recorded under a broad pulse width range of 100 ms - 20 ns. The FTJ measurement protocol for pulsed voltage induced switching is elaborated in the methods section. It is found that in polycrystalline FE thin films, polarization reversal follows a nucleation limited switching (NLS) model, [23] where an independent region-by-region nucleation and domain switching takes place. Switching of the entire system in this case is governed by the domain nucleation statistics with a distribution of mean switching times ($t_{mean}$).

Ferroelectric crystal structure and defects play a significant role in determining domain switching timescales and programming voltages. It has been previously shown that for low thermal budget polycrystalline ferroelectric materials like HZO [35] and P(VDF-TrFE), [23] different post-deposition annealing temperature and protocol can significantly modify the ferroelectric crystal structure leading to different switching dynamics. Also, recent works show that through laser annealing, improvement in crystallization and enhancing $β$-phase content of the P(VDF-TrFE):Graphene composite films is possible. [36] After laser irradiation, typical linear grains of P(VDF-TrFE) was observed, that is quite similar to the 180 °C hotplate-treated film. Both these properties are suitable for improvement of FTJ properties and it's reliability.

Future research will focus on best thermal engineering protocols for improving the dynamic conductance range of FTJs under identical pulse scheme for most efficient training performance.

*2.1.2. Switching timescale and conductance linearity under identical pulse schemes*

Both non-volatile memory and synaptic devices require ultrafast switching. In ferroelectric devices, switching timescale and switching voltage requirement is inversely proportional and therefore it is essential to map their dependence for proper synaptic programming. The resistance switching characteristics of the FTJs, i.e. resistance (*R*) – voltage (*V*) hysteresis loops were measured starting from device off state to on state and back to off state. Programming pulses of a certain pulse width were applied for recording each hysteresis loop. In all cases, one programming pulse is followed by a readout with 0.1 V pulse. The pulse widths were varied between 100 ms to 20 ns range for evaluating the switching speed as a consequence of excitation time. From the *R–V* loops (Fig. 3(a)) obtained by applying rectangular voltage pulses of various pulse widths, the dynamic conductance range of the devices show largest range of few nS to tens of µS while operating with 100 ms pulse widths. With decreasing the pulse widths, the dynamic conductance range starts to decrease and threshold voltages starts to increase for both on to off and reverse directions. In the 1 µs pulse width regime or below, need for higher magnitude programming and erasing pulse was necessary to switch the polarization direction. However even with larger programming and erasing voltages, the polarization switching shows incomplete reversal resulting in much reduced dynamic conductance range. Therefore, shorter pulses in these samples are able to program the devices to intermediate conductance states.

Efficient network training in analog neuromorphic architectures require linearity in synaptic weight update upon application of programming pulses of identical magnitude and width. Previous results demonstrated improvement in synaptic weight update linearity at the expense of increasing the complexity of the synaptic structures, for example, by using a one-transistor/1-resistor (1T1R) or one-transistor/2-resistor (1T2R) structures. However, this results in increased synaptic area overhead and design complexity because of the additional transistors and resistors. [37] Alternatively, higher linearity in 3-terminal synaptic transistors with ionic conduction has been shown. [38] However, these devices show comparatively slower response as their main conduction mechanism is ionic movement driven. Additionally, scalability of ionic transistors are a concern. In our FTJs, linearity of conductance at various pulse widths were tested at different initial conductance states. Broader memory windows at narrower pulse widths are conducive for linear conductance modulation as gradual polarization reversal can be achieved in such junction. In our previous work, [23] we have shown how a deviation from linearity becomes obvious for 100 ns pulses or higher, while conductance change is more linear with 20 ns pulse widths.

One-to-one correspondence of FTJ conductance and ferroelectric domain switching has been shown previously through simultaneous conductance and PFM measurements in oxide perovskite based FTJs [17, 25] confirming that ferroelectric domain switching is the reason for continuous conductance modulation in perovskite oxide based ferroelectric tunnel junctions. Quantitative analysis of domain switching in our ultrathin FE films are performed by fitting the experimental data with theoretical models. A Lorentzian distribution of the logarithm of nucleation times for each applied voltage $V$ with distribution width of $\Gamma(V)$ centered at $\log(t_{mean}(V))$ and the normalized portion of reversed area, $S$ can be calculated as a function of time $t$ and voltage $V$ as,

$$S_{\pm}(t, V) = \frac{1}{2} \mp \arctan \frac{\log(t_{mean}(V)) - \log(t)}{\Gamma(V)} \qquad (1)$$

FE domain configurations and the FTJ resistance $R$ has a one-to-one correspondence, i.e., the fraction of domains with upward ($S$) and downward polarization ($1-S$) can be approximated using a simple parallel circuit model, [17, 39]

$$\frac{1}{R} = \frac{1-S}{R_{On}} + \frac{S}{R_{Off}} \qquad (2)$$

According to this model, the lowest resistance ($R_{On}$) and the highest resistance ($R_{Off}$) can be represented as fully downward ($S = 0$) and fully upward ($S = 1$) switched FE states, respectively. From the experimental conductance ($G$) values after application of a voltage pulse of certain duration ($w$), normalized switched area ($S$) for a certain $w$ were estimated. From the fitting of the $S$ vs. $w$ plots in Fig. 3(b) with Eq. (1), it can be seen that the plots follow the NLS model only for higher applied voltage (-3.5V to 6V). Our previous data confirm this is a feature for the polycrystalline ferroelectric based FTJs with large number of domain pinning centers in the FE film.[23] From the fitting of Eqn. (1), switching parameters like activation energies for FE polarization reversal and corresponding $t_{mean}$ were extracted. The Lorentzian distribution of the logarithm of nucleation times for each applied voltage having width $\Gamma(V)$, is simulated and plotted in Fig. 3(c). The Lorentzian curves centered at $\log(t_{mean}(V))$, shows with lowering of $E$, $t_{mean}$ increases with a broader distribution of switching times. $t_{mean}$ is found to depend exponentially on the electric field ($E$) (Figure 3(d)) as described by the Merz law.[40]

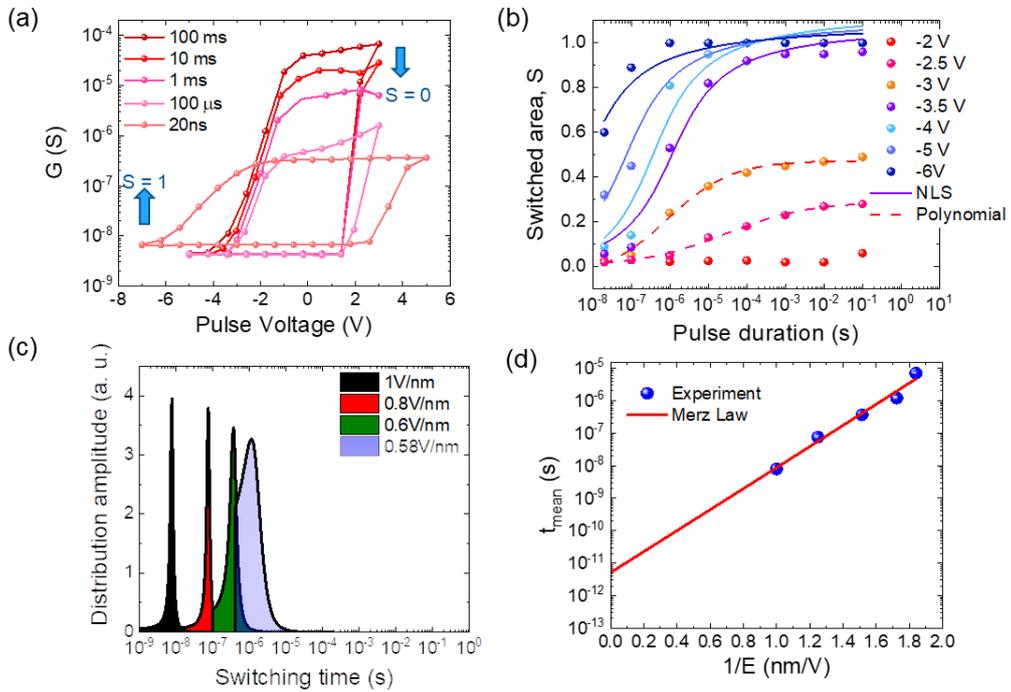

**Figure 3.** (a) Conductance ($G$) hysteresis curves of FTJs for voltage sweeping with pulse durations varying from 100 ms down to 20 ns. (b) Normalized switched ($S$) area as a function of programming pulse duration calculated from $G - V_P$ loops of the FTJs at different pulse

amplitudes and fitting with NLS models. (c) Extracted mean switching times for different electric field values and its distribution around the mean switching times. (d) Evolution of the switching time ($t_{mean}$) as a function of the inverse of the electric field ($1/E$) obtained from fits of the transport data.

Any structural defects or crystalline irregularities in the FE films can result in random domain pinning sites resulting in distribution of $E_C$ and slower or incomplete rotation of FE polarization leading to more gradual change in device conductance. [23] Due to irregular domain pinning centers, higher write noise can also occur due to degree of variability in switching voltages and times over several consecutive programming cycles.

*2.1.3. Optimized pulsing scheme for write linearity*

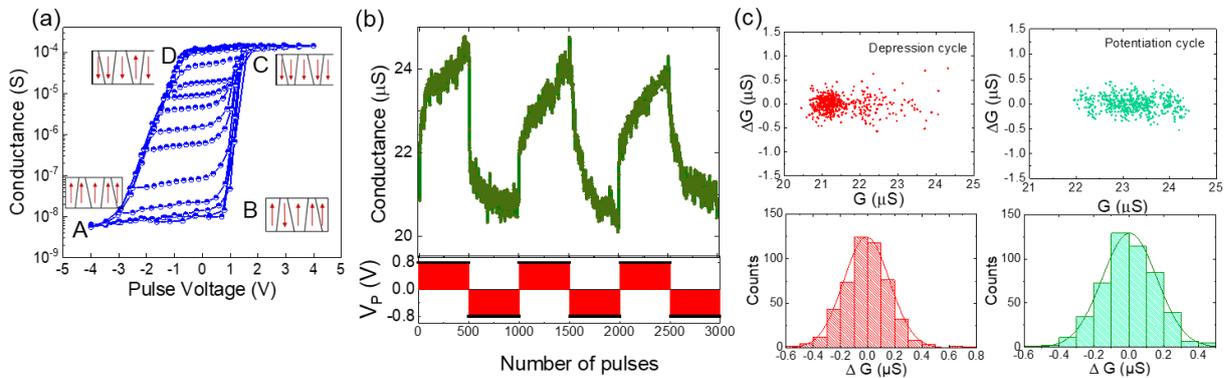

**Figure 4.** (a) Various conductance states of the FTJ memristors when the applied voltages are swept between different ranges. In the present case, a programming voltage pulse ($V_P$) is applied to write a state followed by reading with a 0.1V reading pulse. Arrows indicate the FE polarization direction. Fully up states at point A represent lowest $G$ states while fully down states (point C) represent highest $G$ states. Intermediate states are mixtures of up and down states as shown at points B and D. (b) Conductance potentiation and depression cycles of the FTJs under the application of identical pulse trains of magnitude +0.8V and -0.8V, measured closed to point D. (c) Distribution of $\Delta G$ as a function of $G$ for typical depression and potentiation cycles and the statistics on distribution of $\Delta G$ in each cycle showing the conductance variability that the training operation needs to accommodate.

Fig. 4(a) shows the large $G$ range the device can operate at when operated between ±4V. With lowering the sweeping voltage range, $G$ range starts to decrease due to incomplete polarization reversal. Now, for accelerating the training of DNN, resistive switching memories are expected to satisfy certain conditions that include ± 1V, 1 ns potentiation and depression programming pulses, a symmetric and linear conductance response with ≥32 conductance states (≥5-bit), and a maximum $G$ ($G_{max}$)/minimum $G$ ($G_{min}$) ratio of >10. [13] In FTJ devices with NLS domain switching characteristics, it is challenging to reproducibly obtain large dynamic conductance range ($G_{max}/G_{min}$) with 1 V and 1 ns programming pulses. Based on our evaluation of switched number of domains due to application of voltage pulse of certain magnitude and $w$, (Figure 3), estimations for pulsed parameters were made. Different $G$ ranges were tested in our experiments by setting different initial conductance values. The general observed feature is with increased pulse amplitude or width, a $G_{max}/G_{min}$ ratio >10 can be obtained, however, at the cost of losing linearity and symmetry of weight update as a higher electric field or a longer pulse time results in several domains to switch instantly followed by lesser domain switching after a few pulses. This causes an exponential nature of conductance update instead of a linear one. In the current work, we present one certain pulsing scheme (±0.8 V, 100 ms) related weight update close to point D of Figure 4(a) and the effect of such nonlinearity and asymmetry on the DNN training performance. For these measurements, FTJs were first set to fully down polarized state (point C) and then, alternating pulse trains of 500 pulses of +0.8V and 500 pulses of -0.8V was applied.

Control of domain dynamics is an essential criterion for linear weight update in FTJs and a proper choice of starting conductance value is important for obtaining this linearity. Our choice of point D in the conductance curve for training can be motivated in the following way. In the FTJs with one semiconducting electrode, a transition from off to on state is more abrupt compared to the reverse transition. This points to the fact that rotation of polarization direction away from the semiconductor and hence the creation of a schottky barrier at the semiconductor-FE interface is a more gradual process compared to the elimination of the barrier. Therefore, it is possible to have more controlled conductance modulation in this region through partial rotation of polarization. Distribution of $\Delta G$ as a function of $G$ for a typical depression and potentiation cycles (from Figure 4(b)) are shown in Figure 4(c) and the statistics on distribution of $\Delta G$ in each cycle is shown. The conductance update variability over many measured cycles showed almost similar pattern. Abruptness of domain switching, an inherent property arising from NLS model in polycrystalline FE films, explain the noise margin of the conductance states that could affect the training accuracy of the circuits. However, in our experiments, the network training operation takes place based on weight update over several tens of potentiation and depression cycles, and hence the read noise does not ultimately affect the training accuracies significantly.

In order to quantify the effect of write nonlinearity and noise of the non-ideal FTJs on the accuracy of neural training algorithm, we performed simulations using a modified version of Sandia CrossSim simulator on the MNIST dataset of small and large dimensions (64×36×10 and 784×300×10, respectively). [41] The neural network code is based on backpropagation (BP) algorithm and we used our measured FTJ potentiation and depression properties for the experimental lookup table generation in the simulation. BP is one of the most numerically demanding and energy inefficient neuromorphic algorithms used in a CMOS architecture. Therefore, running BP algorithm on a neural architecture provide a pathway for performance acceleration in terms of energy and computational efficiency.

In our simulation, the FTJ crossbars are used as part of a neural core, i.e., a single layer of the DNN that accelerate the main matrix operations while a digital core is used to process the inputs and outputs to the crossbar, as illustrated in Figure 5.

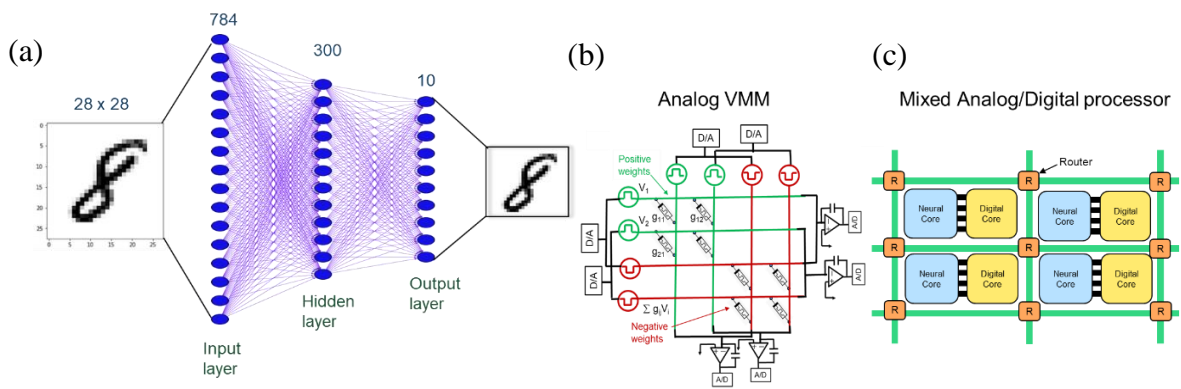

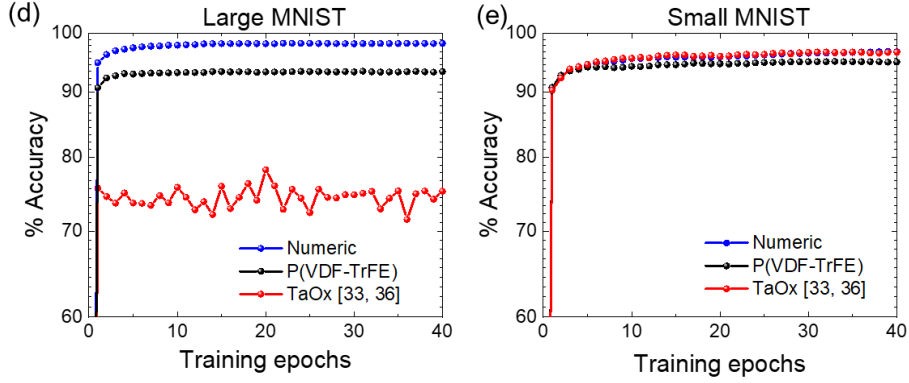

**Figure 5**. (a) MNIST handwritten digit recognition using a 3 layer DNN containing one input, one output and one hidden layer. (b) Memristive crossbar array with FTJs at each crosspoint between the rows and columns where VMM operation takes place though Ohm's law and Kichoff's current law. (c) The neural network accelerator schematics using the neural core as the analog computation engine and digital core for bit quantization, A/D, D/A conversion etc. Training accuracy of the FTJ based DNNs on (d) large MNIST dataset and (e) small MNIST dataset. For benchmarking the system, we also plotted numerical simulation data and one with TaOx memristor data, taken from ref. [41 and 44].

The MNIST training dataset of handwritten single digit numbers contain 60,000 28×28 pixel images of the handwritten single digit of "0" through "9" and additional 10,000 images for testing while for small images, training examples are 3,823 and test examples are 1,797. [42] When using a standard double-precision CPU or GPU to train, a training accuracy of 98% can be achieved, providing a comparison baseline. The simulated training was performed on a network with one input, one hidden and one output layer (Figure 5(a)). In order to simulate the response of our FTJ device in a neuromorphic algorithm, a lookup table is created from the measured potentiation and depression data in response to potentiating and depressing pulses of identical magnitude and width. $\Delta G/G$ was extracted from the cycling data together with the probabilistic change in conductance at a given $G$ value. This $\Delta G/G$ values and the cumulative distribution function (CDF) are then used to predict how a device will respond while training.

In our simulation, BP algorithm have been implemented in the following way. Each layer of the DNN has been represented by a simulated crossbar of FTJs to perform two key operations, a vector matrix multiply (VMM), and parallel outer product update. All FTJ memristors in the crossbar array are updated in a single step. This leads to efficient parallel updates that improves energy efficiency and latency. Here, the approximation is the magnitude of the weight update is linear in time, current and voltage, *i.e.* by linearly varying time, current or voltage, we can linearly modify the conductance change, *ΔG*. The linearity is based on the measured potentiation and depression linearity, as shown in Figure 4(b). The analog core, here, performs the VMM operation by utilizing the Ohm's law and Kirchoff's current law (Figure 5(b)). Voltage pulses are fed to the rows of the crossbar and current output from the column of the crossbar is extracted. We used sigmoid neurons with a slope of 1 and a learning rate of 0.1 that can perform classification tasks with a high level of accuracy. In BP, after each training cycle, the actual outputs of the computing circuits are compared with the expected output, and the difference (*i.e.*, the error) is fed back *i.e.* backpropagated to make a correction on the weights, *G*.

**2.2. DNN training with FTJ as programmable synaptic weight elements**

*2.2.1. Training accuracy based on conductance range*

As the physical memristor components possess a $G_{max}$ and $G_{min}$ in response to a train of potentiating and depressing pulses, the network has a minimum and maximum weights it can store. Choosing the correct weight range, therefore, ensures optimized usage of the dynamic conductance range of the memristors, with minimized effect of nonlinearity. We investigated the effect of mapping the weights at the center 50% and towards the edges of the *G* range of the values shown in figure 4(b). Also the chosen *G* range for different pulsing schemes based potentiating and depression cycles were tested. Based on the linearity and symmetry of the *G* updates, improvement of the accuracy is obtained for certain chosen ranges, improving mainly in the region where average conductance change, *ΔG* is nearly constant. During training cycles, some weights went outside the targeted center 50% range, however, majority of the weights remained in the targeted range.

In FTJs, continuous pulsing with subthreshold voltage (so that $E<E_C$) leads to linear analog variation in $G$ due to gradual rotation of FE domains. However, the domain rotation in FE materials depends on many factors like structural defects acting as domain pinning centers, size of the domains, etc. For ultrathin FE films, additional factors like ratio between thickness and area and interface induced stress can also be decisive. Therefore, with continuous pulsing, we often end up having an exponential $G$ update, i.e. changing quickly initially and moving towards saturation gradually. For analog VMM operation, however, it is important to choose a conductance range where $G$ varies almost linearly. Based on the chosen $G$ range, the training accuracy of the network varies. The chosen $G$ range does not have that significant impact on the training accuracy for the small images, however, for the large images at least a difference of 2% was visible. This finding opens an avenue for investigating effect of engineered domain pinning sites on weight update linearity and symmetry for neural accelerator training.

Network training applications generally do not require the 10-year retention period needed by non-volatile memories like NAND flash, and often target training periods of less than one week after which the weights can be read and stored digitally, if needed. For on-line training, a high retention period is, therefore, not essential. However, for an inference only application, it is a desired quality, otherwise the inference accuracy suffers. If the memory retention is low, the trained weights need to be transferred to a non-volatile memory such as flash before the inference starts. This increases the energy consumption and latency of the system. As our FTJs can retain the programmed state for substantially long time, [20,22] systems designed with FTJs are expected to operate without frequent data transfer, increasing the system efficiency.

*2.2.2. Training accuracy based on applied pulse width*

Proper control of FE domain dynamics through custom-designed pulse amplitude and width can result in fine-resolution programmable memristive functions for various circuit applications. In our experiments, we found that shorter pulses of 20 ns pulse width leads to better $G$ update linearity [23] and can improve the training accuracy. This result however, requires, programming pulses of higher magnitude and larger asymmetry between positive and negative pulses and therefore from circuit implementation point of view, certain complexity can arise. This is currently under study and will be communicated in future reports.

*2.2.3. Training Accuracy based on peripheral circuit parameters*

As mentioned previously, the analog crossbars act as a neural core in our experiments, while a digital core is used to process the inputs and outputs to the crossbar. Here, we used finite analog-to-digital (A/D) and digital-to-analog (D/A) precisions. Depending on the specifications of the digital circuit bit precision, the network training accuracy was significantly modified. For instance, reducing the A/D and D/A bit precision from 8 bits to 4 bits reduces the training accuracy from 93% to 84% for large digit recognition while that for small digit recognition changes from 95% to 93%. In our simulation, we did not take into account the impact of circuit parasitics and access devices, as the sole focus of the current work is to test the training accuracy limitations imposed by the FTJ device non-idealities, when operated under identical pulses. However, design optimization for circuit parasitics can be made for more realistic outcomes. [43] Additionally, as the FTJs have resistance that is orders of magnitude higher compared to the line resistance, current through the leakage paths can be considered negligible. In FTJs with a semiconductor electrode, each device can represent one diode and one memristor itself reducing leakage issues and can lower circuit overheads. However, large-scale experimental integration is needed to confirm this hypothesis.

*2.2.4. Training accuracy based on network size*

Number of parameters involved in computations in neural networks scale up drastically with the size of the network and can impose a huge burden on hardware resources, limiting the processing speed, and consuming huge amount of energy. In our training simulation, we used two different sizes of the network to have an estimation on the device non-linearity and asymmetry on the size of network. As can be seen from Fig. 5(d) and (e), *G* nonlinearity and asymmetry starts to play a significant role as the size of the network increases. For a nonlinear and asymmetric memristor, like TaOx RRAMs, [41, 44] the difference in training accuracy between small and large image training is almost 20%. For our FTJ memristors, a 2% difference in training accuracy is obtained for small and large handwritten datasets. While changing the *G* range, or A/D and D/A quantization parameters, larger networks were found to have significant impact compared to the smaller ones. The CPU run time on an average for large images is 1.3 hours per training and testing epoch while that for the small images is 4.8 seconds.

Another point of concern in FTJ based large-scale arrays is the large *RC* constants that could lead to non-trivial sensing delays. Luo et al. [45] showed through simulation that in HZO based FTJ crossbar arrays, the sensing delay for the $1024 \times 1024$ array is about 40 ns. Sensing delay does not change much with respect to the array size when the array is small (N < 256), while sensing delay grows linearly with the array size when the array is large (N > 256). Additionally, device to device variation and fabrication related defects could cause predicted training accuracy to deviate significantly. Array level experimental result of FTJ crossbars and further upgradation of the simulation platform will lead towards a more realistic result and future research will concentrate on this topic.

*2.2.5. Device level benchmarking*

In the Table 1, the key performance matrices of ferroelectric two and three-terminal memory devices are listed for their implementation in neuromorphic computing hardware. As the hardware implementation of ferroelectric memristor based neural accelerators would require process parameters compatible with CMOS back-end-of-line thermal limit, we restrict our discussion to low thermal budget ferroelectric devices only. Polymer FEs can be processed at <150 $^0$C and therefore they can be integrated on flexible and wearable devices as well.

**Table 1. Bechmarking ferroelectric memory devices**

| FE Material | Technology | On/Off ratio | Write voltage | Write time | Retention (s) | Endurance (cycles) | Process temperature | Ref. |
|---|---|---|---|---|---|---|---|---|
| HZO | FeFET | $10^6$ | 3V | <100 ns | $>10^4$ s | $>10^{10}$ | 500 $^0$C | 14 |
| HZO | FeFET | $10^6$ | 2V | <1 µs | $>10^8$ s | $>10^8$ | 500 $^0$C | 46 |
| HZO | FeRAM | $10^8$ | 2.5V | <10 ns | $>10^3$ s | $>10^{11}$ | 500 $^0$C | 47 |
| HZO | FTJ | 30 | 3V | - | $>10^5$ s | - | 500 $^0$C | 48 |
| HZO | FeFinFET | $>10^4$ | 3.5 V | 100 ns | $>10^8$ s | $>10^{11}$ | > 500 $^0$C | 15 |
| P(VDF-TrFE) | FeTFT | $10^4$ | 8V | <1 ms | $>3 \times 10^3$ s | $>10^7$ | 200 $^0$C | 49 |
| P(VDF-TrFE) | FeRAM | $10^5$ | 4V | - | $>10^3$ s | - | 170 $^0$C | 50 |
| P(VDF-TrFE) | FTJ | $>10^4$ | <5V | <20 ns | $>10^4$ s | $>10^3$ | 140 $^0$C | 23 |
| P(VDF-TrFE) | FTJ **This work** | $>10^4$ | 0.8V | 100 ms | - | - | 130 $^0$C | - |

Based on different application domains, computing circuit paradigms, size of the network, training protocols and algorithms etc., the training accuracy can change significantly. For instance, Agarwal et al. [51] showed using an additional "periodic carry" method in BP algorithm an RRAM based DNN accelerator's accuracy can be improved from 80% to >97%. At the same time, a linearly increasing programming pulse magnitude and pulse width has been shown to improve linearity of weight update over constant magnitude and width pulse trains, thus improving training accuracy. [13] However, this improvement comes with the cost of circuit complexity. Therefore, doing a one-to-one comparison of the training accuracy is not preferable unless all training parameters and algorithms are similar. Still, for the sake of highlighting some recent works on ferroelectric NVM based neural network training, a benchmarking is done in Table 2 where multi-layer perceptron (MLP) based network training simulation is performed on MNIST handwritten dataset.

**Table 2. Bechmarking training accuracy**

| FE Material | Technology | $G_{max}/G_{min}$ | Pulse Scheme | Maximum V | $R_{On}$ (Ω) | Number of states | Training accuracy | Ref. |
|---|---|---|---|---|---|---|---|---|
| HZO | FeFET | 45 | Incremental | 3.65, 75 ns | 500 k | 32 | 90% | 13 |
| HZO | FeFinFET | 45 | Incremental | 8V, 100 ns | 10 M | 27 | 97% | 15 |
| HZO | FTJ | 2 | Identical | 3V, 100 ns | 10 k | 10 | 96% | 52 |
| HZO | FeTFT | 100 | Incremental | 6V, 100 ns | 100 k | 8 | 95% | 53 |
| HZO | FeTFT | 14 | Incremental | 4.3V, 10 ms | 1 M | 60 | 91% | 30 |
| P(VDF-TrFE) | FTJ **This work** | 1.2 | Identical | 0.8V, 100 ms | 10 k | > 8 | 93-95% | - |

## 3. Conclusion

In conclusion, we have reported how precise control of FE domain dynamics can lead to efficient DNN training. An integrated device-to-algorithm framework for benchmarking novel synaptic devices is used for showing DNN training based on polycrystalline FE based MFS tunnel junctions. Training accuracy from the system is comparable to the values reported from perovskite FTJs and HZO based FTJ, FeTFT and FeFETs, however, using identical pulsing schemes. Possibility of precise domain control in these FTJs with proper choice of pulsing scheme and engineered domain pinning sites can lead to better weight update linearity and symmetry of the FTJ synapses and can result in improved DNN accelerators. Rather limited dynamic range for pulsing with ±0.8V is still a challenge. However, reproducible intermediate states over repeated potentiation and depression cycles with small amount of noise offer possibilities for multibit operation. Innovative architecture and device stack design could improve dynamic ranges and is under current investigation.

## 4. Experimental Methods

*Device Fabrication:* The FTJs were prepared by spin coating P(VDF-TrFE) thin films onto NSTO substrates with Nb concentrations of 1 wt%, obtained from Surfacenet GmbH. The as-received substrates were thoroughly cleaned in ultrasonic bath with Acetone, Isopropyl alcohol and DI water followed by drying with $N_2$. For spin coating, we prepared a P(VDF-TrFE) solution by dissolving 70:30 copolymer powder from Piezotech in methyl-ethyl ketone (MEK) with a solution strength of 0.2%. The solution was spin-coated at 4000 rpm for 1 minute followed by annealing in air for two hours at 130 °C with a slow cooling to room temperature. This process resulted in 6-nm-thick P(VDF-TrFE) ferroelectric films, as confirmed by atomic force microscopy (AFM) measurements. [20] Finally, we used electron-beam evaporation to grow 100-nm-thick Au electrodes through a metal shadow mask. The size of the top electrodes was 300 x 300 µm$^2$.

*Device Characterization:* Atomic Force Microscopy (AFM) and Kelvin-probe force microscopy (KPFM) measurements were conducted in a Bruker Dimension 5000 system in tapping mode using Pt/Ir-coated Si cantilever tips. The thickness of the P(VDF-TrFE) films was calibrated by performing AFM scans on prepared step edges. For local PFM, an ac voltage of amplitude 1 V was superimposed onto a sweeping bias voltage. The voltages were applied to the cantilever tip while the bottom electrode was grounded. We operated the PFM system near a resonance frequency of ~350 kHz. Electrical transport measurements on the FTJs were performed at room temperature. We used a probe station to contact the bottom and top electrodes in a two-point measurement geometry to a Keithley 2400 sourcemeter or Keithley 4200 semiconductor analyzer. Voltage pulses were applied to the junctions using an arbitrary function generator (Tektronix AFG 1062). In all measurements, the NSTO bottom electrode was grounded and voltages were applied to the Au top contacts. For measurements shown in Fig. 3(a), each conductance loop corresponds to negative to positive and reverse sweep of voltage pulses of varying duration. Pulses are applied to the FTJ's top electrode (keeping the bottom electrode grounded) and the junction resistance is measured at a $V_{read} = 0.1$ V after each write voltage pulse. For training experiments, conductance potentiation and depression cycles of the FTJs were measured under identical pulse trains of magnitude +0.8V and -0.8V. One write pulse of 0.8V and one read pulse of 0.1V were followed by the next set of write and read pulses. For All measurements are performed at room temperature and at ambience.

*Modelling and Deep Neural Network Simulation:* Modelling of FTJ device experimental data was performed with codes written in MS-Origin program while DNN training simulation experiments were performed using a Python code that is a modified version of the Crosssim simulation platform.

**Acknowledgements**

The author thanks the Academy of Finland (Grant no. 345068 and 350667) for financial support. Dr. Binbin Chen is acknowledged for performing the KPFM and PFM measurements of the P(VDF-TrFE) films. The work used experimental facilities of Micronova national research infrastructure for micro- and nanotechnology.